%% file: main.tex
\documentclass[sigconf]{acmart}
\usepackage{graphicx}
\usepackage{graphicx}
\usepackage{subcaption}
\usepackage{comment}
\usepackage{caption}
\usepackage{siunitx}
\usepackage{booktabs}
\usepackage{pifont}
\usepackage{hyphenat}
\usepackage{algorithm}
\usepackage{algorithmic}

\graphicspath{{./images/}}
\usepackage{pdfpages}
\usepackage{bytefield}

\setcopyright{acmcopyright}
\copyrightyear{2021}
\acmYear{2021}
\acmDOI{xx}

\acmConference[draft]{draft }{Jan 07--11, 2022}{Shanghai,China}
\acmBooktitle{draft}
\acmPrice{15.00}
\acmISBN{XXX-X-XXXX-XXXX-X/XX/XX}

\begin{document}

\title{Ruta: Dis-aggregated routing system over multi-cloud }

\author{Kevin Fang}
\affiliation{%
  \institution{Cisco Research \& Development Center }
  \city{Shanghai}
  \country{China}
}
\email{zhiyfang@cisco.com}

\input{sections/abstract}

\vspace{1em}

\maketitle

\input{sections/introduction}

\input{sections/controlplane}

\input{sections/dataplane}
\input{sections/system}
\input{sections/conclusion}
\input{sections/acknowledgment}

\bibliographystyle{ACM-Reference-Format}
\bibliography{references.bib}

\end{document}

%% file: sections/abstract.tex
\begin{abstract}

Over the years, the SDN evolution create multiple overlay technologies which is inefficient and hard to deploy end-to-end traffic engineering services, Ruta is designed as an unified encapsulation with Segment Routing, Crypto and NAT-Traversal capabilities over UDP. 

Ruta could be deployed as a cloud native SDN platform globally over multi-cloud and integrated with each applications on transport layer, which provide \emph{nearly zero loss and almost less than 200ms latency to access anywhere in the world over internet}.

\end{abstract}

\ccsdesc[500]{Computer systems organization~Cloud computing}
\ccsdesc[500]{Networks~Transport protocols}
\ccsdesc[500]{Hardware~Networking hardware}

\keywords{Routing, Overlay Network, Software Defined Network, Distributed Controllers}

%% file: sections/introduction.tex
\section{Introduction}
\label{sec:introduction}

Software Defined Network and Cloud VPC evolution create multiple overlay technologies in last decade, the different implementation separate network into multiple domains. Meanwhile applications require simplicity for end-to-end traffic engineering and security policy enforcement, the domain-specific SDN design cause many overheads on both control and data plane.

\noindent \textbf{Paper Organization:} We introduce the challenges of overlay technologies used today and the motivation for Ruta project in  Section~\ref{sec:introduction}. We present the control plane architecture in Section~\ref{sec:controlplane}, data plane architecture in Section~\ref{sec:dataplane} and prototype system implementation and demonstrate multi-cloud deployment in Section~\ref{sec:system}.We conclude in Section~\ref{sec:conclusion}.

\subsection{Overlay Technologies}

A packet from client to cloud require multiple times encap/decap and encryption/decryption which introduce significant latency and inter-working complexity on each boarder network devices . 

A packet which send by a wireless client require \emph{location-agnostic}, vendors like Cisco~\cite{ciscodnac} or juniper~\cite{junipermist} implement overlay and group based policy for wireless converged campus network. Each vendor has private control-plane protocol(lisp~\cite{lisp}, BGP-EVPN) and data-plane( vxlan-gbp~\cite{vxlangbp} vxlan-gpe-gbp and lisp-gpe-gbp ~\cite{vxlan_gpe_gbp}). The campus network fabric policy design based on \emph{user-centric} approach, thus group based policy tag only contain user or group identity.

When the packet arrive campus boarder, SDWAN Router will decap the packet and enable \emph{deep-packet inspection(DPI)} for application aware routing or security policy enforcement, packet will be encap and encrypted in IPSec or DTLS with a transport VPN tag, then sent to remote on-prem datacenter or public cloud. In SDWAN domain policy design based on \emph{transport-centric} approach. Most of the SDWAN implementations are based on point-to-point tunnel,thus the multi-hop traffic engineering may require multiple times encryption and decryption with multiple times policy lookup. Some of the service provider use MPLS-VPN, MPLS-Segment Routing or SRv6 as overlay to handle the traffic, MPLS require dedicated underlay and SRv6 require IPv6 link, and the SRv6 forwarding mechanism does not change the source address which may cause problems when unicast reverse path forwarding(uRPF) enabled, so most of them can not directly transport over internet. Even with segment routing over IP (RFC8663) may support MPLS-SR over IPv4 UDP, but the forwarding mechanism does not support NAT-Traversal.

When packet arrive on-prem datacenter, the network policy design based on \emph{application-centric} approach, group based tag like Endpoint Group(EPG) in Cisco ACI architecture~\cite{ciscoaci} identify application sever groups. The boarder gateway have to update the source and lookup the destination group id and set related VNID and group ID. 

It's more complex in public cloud, Virtual Private Cloud(VPC) are based on region and available zones, inter-VPC communication require complex routing and security group policy which requires VPC-Peering or Transit-VPC services. \emph{cloud-agnostic} require more generic overlay network support, but different cloud provider may have difference overlay terminology introduce significant workload for multi-cloud deployment.

\subsection{Design principal}

Inspired by cloud native approach, we design \textbf{Ruta} based on service-mesh architecture, we dis-aggregate routing service to distributed K-V control-plane and micro-service based data-plane to provide unified encapsulation to meet each network domain's requirement. It could significantly reduce the inter-working complexity and provide flexible programming interface for applications. 

However, Design dis-aggregated routing platform is not simply leverage distributed software architecture to build controller and data-plane, but it require more carefully trade-off on architecture level,because the system failure may easily cause network partitioning and lose availability.
Based on CAP theorem ~\cite{cap}, some design principals are shown as below.

\emph{Consistency} is the must to have feature in traditional network,the destination-based forwarding requires all nodes have \emph{consistent} forwarding table, Inconsistent forwarding table scenario may well-known as “Micro Loop” or “Black hole”.  

Meanwhile, the traditional routing protocols are designed to serve destination based forwarding. No matter which routing protocol were used, it must implement a routing information database(RIB) and keep \emph{availability} for data-plane calculate the forwarding information database(RIB). Most of the routing protocols soften the partition tolerance as a trade-off, for instance OSPF-area or BGP-Autonomous System are designed to hide topology and isolate failure.

At the same time, keeping availability may loss \emph{partition tolerance}, many routing protocol need to deal with "brain-split" situation. Leader arbitration, split-horizon are designed to reduce the network partition impact, but it has side effect for the availability.

In the SDN Era, control-plane is centralized, many network devices need sequential consistency, and the DHCP like address assignment need strictly consistency. The controller implementation and placement under network partition becomes major challenge. For instance, recently facebook outage~\cite{fboutage} indicate the controller availability is the root cause.

In summary, the consistency requirement and destination based forwarding is a kind of trade-off few decades ago. In the old days, the forwarding ASICs or Network Processors may have limited forwarding capabilities and the network has limited bandwidth to carry more information for source routing. 

Can we soften the \emph{consistency} requirement by introducing the source-routing or segment-routing(SR) and decouple the devices configuration ?

Can we soften the \emph{availability} requirement by using distributed path-compute on each forwarding adjacency when running into the headless mode?

\begin{figure}[h]
	\centering
	\includegraphics[scale=0.45]{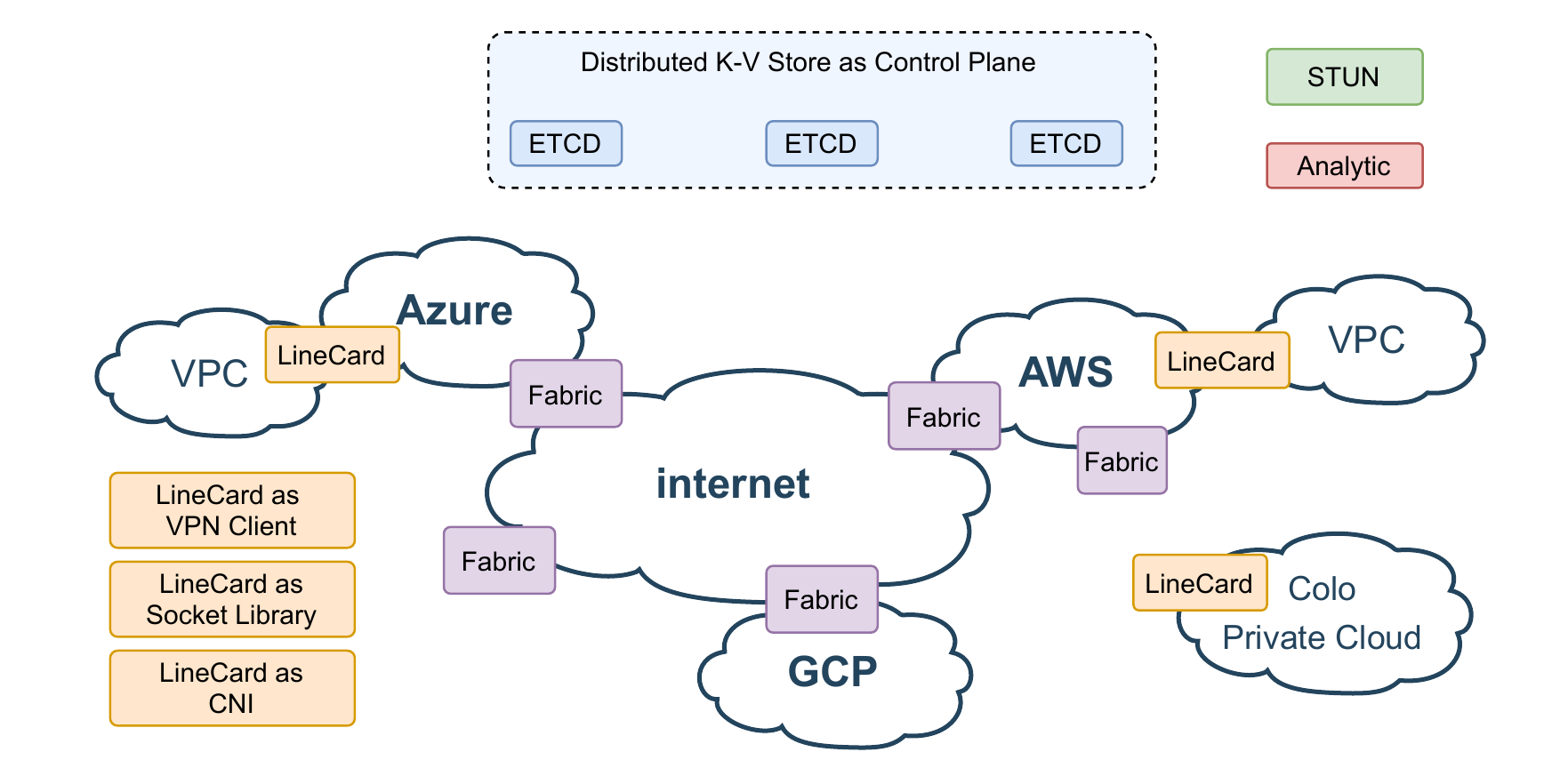}
	\caption{Ruta: Dis-aggregated routing architecture}
	\label{Fig:ruta_archi}
\end{figure}

These are the design principals for \textbf{Ruta} system, we implement an \emph{eventually consistency} model to decouple the prefix announcement and path-computation process by using a location descriptor, then we leverage local cache mechanism for distributed path-computation during control-plane offline. We select ETCD as our control-plane and simplify all routing updates and decouple the policy related configurations in K-V pairs. 

Ruta separate the consistency requirement by using distributed path-calculation and segment routing to decouple overlay route information and linkstate, linecard could use partially linkstate db for path calculation without global consistency requirement, In some large scale deployment, users could deploy multiple redis based service node for regional linkstate database and provide cache service for linecards.

Segment Routing(SR) with loop-free alternative provide better resilience under network partition, however it only support IPv6 or MPLS, even RFC8663 may support MPLS based SR over UDP, but it lack of programmability and NAT-traversal support. Meanwhile,as a cloud-native routing system, we need to use a pure user-space dataplane to bypass kernel overhead and simplify the programmability for application developers. SRv6 may require kernel support to add IPv6 Option header which is not friendly for application developers. 

We develop a new data-plane protocol by using segment routing over UDP(SRoU). SRoU header can be integrated with QUIC to provide secure and reliable transport services. Many public cloud and many internet service still using IPv4 and deal with the SRv6 uRPF issue, we encode the original source address in SRoU, then we develop STUN services to resolve IPv4 NAT-Traversal problem.


%% file: sections/controlplane.tex
\section{Control Plane}
\label{sec:controlplane}

Ruta control plane is heavily use the ETCD features like lease, distributed lock and transparent proxy. Each service node could be a proxy for others connect to ETCD. All communications to ETCD require TLS, RBAC could be added enhance system security.

\subsection{Service Node Roles}
\label{subsec:snode_role}
Ruta define 5 roles of service node which is shown in Figure.\ref{Fig:ruta_archi}.

\begin{itemize}
	
	\item\textbf{ETCD:} This node is running ETCD process to provide distributed K-V store services as Ruta control-plane. All Ruta node info, routing table, policy and link state are represent in K-V pairs.
	
	\item\textbf{Fabric:} This node type is used as an middle box to relay the SRoU packets. This node could be implemented by in DPDK based appliance, or offload packet processing on DPU or pure P4 based switch to provide high bandwidth forwarding capabilities.  This node must enable link probe to other Fabric node and report link-state to K-V store.
	
	\item\textbf{Linecard:} This node fetch routing table and topology from ETCD and execute flexible algorithm to find available path and encode SRoU SID-list in each packet. Linecard node has various types to provide basic transport service or tightly integrate with applications.It could be physical SDWAN routing box , mobile vpn client, sidecar proxy, DPU or even a QUIC socket library integrated with applications.
	
	\item\textbf{STUN:} This node is used as a STUN(Session Traversal Utilities for NAT,RFC5389) server to help IPv4 based Fabric node and linecard node to retrieve their public address and port.
	
	\item\textbf{Analytic:} This node is used to watch stats from ETCD, analysis network failure and gives proactive response for policy change based on AIOps approach.
	
	\item\textbf{LSDB(optional):} This node is used as a regional linkstate database and cache to mitigate ETCD performance issue with in very large-scale and global deployment. This node contains a Redis DB for linkstate information.  Fabric and Linecard node could use geometry information hunting the nearest LSDB Node.
	
\end{itemize}

\subsection{Node Registration}

All service node must register to ETCD based on the following K-V pairs

\begin{itemize}
	
	\item\textbf{Key:=} "/node/{\color{orange}<role>}/{\color{cyan}<systemName>}"
	
	{\color{orange}<role>} is defined in section ~\ref{subsec:snode_role}, {\color{cyan}<systemName>} is a system wide unique value, just like traditional Router-ID, users could continues use router-id in this field,or use hostname string instead.
	
	\item\textbf{Value:} contains \emph{Site-ID}, \emph{Location} and \emph{SystemLabel}
	
	\subitem\textbf{Site-ID:} This field is used for site level policy enforcement.
	\subitem\textbf{Location:} This field store the node latitude and longitude information. Linecard node could apply flexible geo-aware-routing or build random-graph to reduce computational complexity in large scale network.
	\subitem\textbf{SystemLabel:} This is 24bits field for Segment ID compression and MPLS interworking. Label assignment is implemented by distributed lock mechanism over ETCD, each node acquire the lock from ETCD then assign unique smallest Number as SystemLabel, and register it back to ETCD with SystemLabel field.
	
\end{itemize}

\subsection{Service Locator(SLoC)}

Each of service node has multiple interfaces, Ruta use \emph{Service Locator(SLoC)} describe them.

\begin{table}[h!]
	\small
	\begin{tabular}{m{1.6cm}m{5.4cm}}
		\toprule
		\textbf{Field} & \textbf{Usage}\\
		\hline
		Color & This field is used for link level policy enforcement, user may defined it based on link type or coloring different links(same as color definition in SR-Policy).  \\
		\hline
		Private IP\&Port &  The private IP address and UDP port used for Ruta service.\\
		
		\hline     	
		Public IP\&Port &  The public IP address and UDP port used for Ruta service.\\
		
		\hline     	
		Interface Name &   optional field to store the local interface which initial the SRoU session.\\
		\hline     	
		RX/TX BW &  max upstream/downstream bandwidth, used for calculate link utilization.\\
		
		\bottomrule
	\end{tabular}
	\caption{Service Locator(SLoC) Field}
	\label{tbl:SLoc_field}
\end{table}
Ruta may use the shorten format to distinguish interfaces.

$SLoC_{short}$ := {\color{cyan}<systemName>}|{\color{orange}<color>}|<Private IP:Port>

\subsection{Service Hunting}
After Node registration, service node must send \emph{Service Location Route} for service discovery.

\begin{itemize}
	
	\item\textbf{Key:=} "/service/{\color{orange}<role>}/{\color{cyan}<systemName>}"
	
	\item\textbf{Value:} SLoC data structure will be used in this field.
	
\end{itemize}

\noindent Service Node could use prefix based fetch {\color{orange}/service/<role>} for service hunting, For instance, a new on-boarding device need STUN service, it may fetch {\color{orange}/service/STUN} as prefix from ETCD to hunt STUN server's public ip address and port.

\subsection{Link State}

When Fabric node finished the on-boarding process, it must fetch the {\color{orange}/service/lsdb} to discover the LinkState Database node. it must use prefix based fetch {\color{orange}/service/Fabric} to discover other fabric nodes, then start full-mesh link state probe to others. In some large scale deployment, Fabric node could config a white-list to reduce full-mesh probe.

The probe result need to update in LSDB Redis DB, if system does not contain LSDB node, it must send to ETCD, Link state K-V pair is shown as below:

\begin{itemize}
	
	\item\textbf{Key:=} "/stats/{\color{orange}linkstate}/<{\color{cyan}$SLoC_{src}$} - {\color{red}$SLoC_{dst}$}>"
	
	\item\textbf{Value:} Ruta Link state probe leverage the algorithm from Two-Way Active Measurement Protocol(RFC5357)~\cite{twamp}, it could provide two-way delay, jitter, loss measurement, it will also report the link utilization and up/down status.
	
\end{itemize}

\subsection{Service Route}

Inspired by Locator/Identifier Separation Protocol(LISP), Ruta service route does not contain any explicit nexthop but just mapping the uniform resource identifier(URI) to SLoC.
Ruta URI service route framework not only designed for packet routing service, but also support various of new services(eg. multi-cloud RPC, edge computing node).

For multi-cloud packet routing services, Ruta could carry EVPN-Route as resource identifier.

\begin{itemize}
	\item\textbf{Key:=}"/route/{\color{cyan}<type>}/{\color{orange}<export RT>}/{\color{blue}<RD>}/*"
	
	{\color{orange}Route Tagert(RT)} and {\color{blue}Route-Distinguisher(RD)} usage are same as BGP-EVPN, RT mechanism could be implemented by watch "/route/{\color{cyan}<type>}/{\color{orange}<RT>}/" prefix. 
	
	EVPN route URI listed below:
	
	\begin{table}[h!]
		\small
		\begin{tabular}{m{1.5cm}m{5.2cm}}
			\toprule
			\textbf{Type} & \textbf{Key}\\
			\hline
			EVPN Type2 & /route/{\color{cyan}2}/{\color{orange}<export RT>}/{\color{blue}<RD>}/MAC/IP \\
			
			EVPN Type5 & /route/{\color{cyan}5}/{\color{orange}<export RT>}/{\color{blue}<RD>}/IPPrefix/Mask\\
			
			\bottomrule
		\end{tabular}
		\caption{ EVPN Route URI Format}
		\label{tbl:sroute_format}
	\end{table}
	
	\item\textbf{Value:} just like BGP attributes, it has 3 mandatory field(SiteID, SystemName, PolicyTag). TLV based optional field extension will be used in the future.
	
\end{itemize}

\subsection{Path Computation}

Unlike traditional routing protocol, Ruta is used for overlay transportation. The path computation logic is more like google map navigation, it could support various of algorithms to meet difference SLA.Meanwhile it does not require consistency of link-state database which is very useful during network failure.

By default, the linecard will build destination SLoC list based on EVPN routes, it will only use active link state probe for each destination service node. If some of the route has SLA violation, the linecard could use a fabric-node-list or randomly selected fabric nodes to fetch the related link-state, and execute local path computation.

Path Computation engine will generate a SLoC list for each EVPN route. This list will be encoded in SRoU header.

\subsection{Node Keepalive}

Service Node keepalive leverage the ETCD lease function, When a node failed to update lease time, the ETCD will automatic withdraw the related information. Consider the controller availability we implement 2 different lease time. The first lease time(60~120 seconds) is used for service node keepalive. The second lease time(600 seconds ~ 1200seconds) used for linkstate and service route.

\subsection{Micro Segmentation}

\textbf{Endpoint Identity:} Each of the endpoint may have it's identity or group policy tags, it could be updated by the following K-V pairs

	\textbf{Key:=}"/identity/userid/device-id" 
	
	\textbf{Value:=} "group policy tags"

\noindent\textbf{Group based policy:} Each node may use the following K-V pairs for group based micro-segmentation:

\textbf{Key:=} "/control/group/$SRC_{group}$/$Dst_{group}$"

\textbf{Value:=}  "Action"|"SLoC list"

\noindent\textbf{Route control:}Network operator could update the control policy to the entire system by using:

\textbf{Key:=} "/control/RT/2/$SRC_{MAC}$/$SRC_{IP}$/$DST_{MAC}$/$DST_{IP}$"

\textbf{Key:=} "/control/RT/5/$SRC_{prefix}$/$SRC_{mask}$/$DST_{prefix}$/$DST_{mask}$"

\textbf{Value:=}  "Action"|"SLoC list"

%% file: sections/dataplane.tex
\section{Data Plane}
\label{sec:dataplane}

Ruta dataplane leverage the SRv6 programmable SRH concept, but move the SRH after the UDP header, this new encapsulation called SR over UDP(SRoU).It add explicit FlowID field for micro-segmentation. Source IP address and port are added for NAT-Traversal.

\begin{figure}[h]
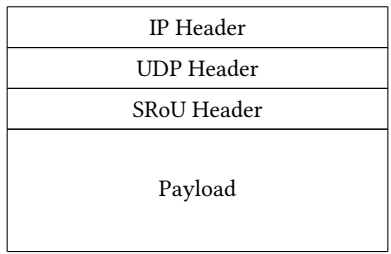

	\centering
\begin{bytefield}[bitwidth=0.5em]{32}
	\wordbox{1}{IP Header} \\
	\wordbox{1}{UDP Header} \\
	\wordbox{1}{SRoU Header} \\
	\wordbox{3}{Payload} \\
\end{bytefield}
	\caption{SRoU Encapsulation}
\label{Fig:srou_encap}
\end{figure}

\subsection{SRoU Header}

SRoU header defined as below:

\begin{figure}[h]
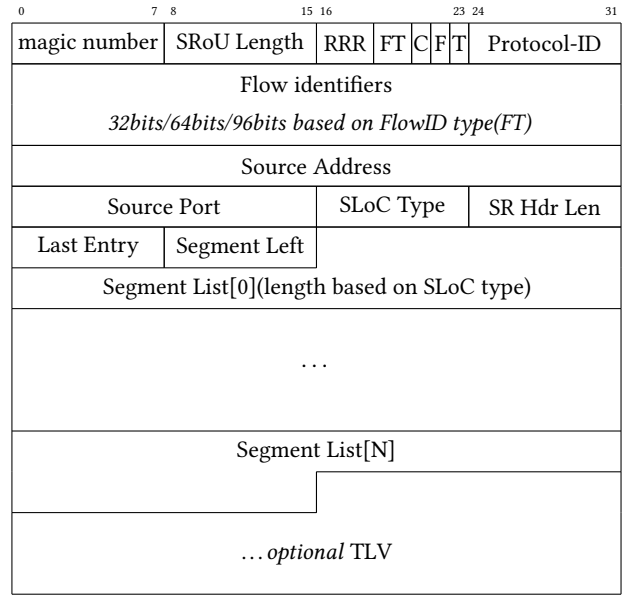

	\centering
\begin{bytefield}[bitwidth=0.8em]{32}
	\bitheader{0,7,8,15,16,23,24,31} \\
    \bitbox{8}{magic number} & \bitbox{8}{SRoU Length} & \bitbox{3}{RRR} \bitbox{2}{FT} \bitbox{1}{C}\bitbox{1}{F}\bitbox{1}{T} &\bitbox{8}{Protocol-ID}\\
	\wordbox[tlr]{1}{Flow identifiers} \\
	\wordbox[blr]{1}{\emph{ 32bits/64bits/96bits based on FlowID type(FT)}} \\
	 \bitbox{32}{Source Address} \\
	 \bitbox{16}{Source Port} & \bitbox{8}{SLoC Type} &\bitbox{8}{SR Hdr Len}\\
     \bitbox{8}{Last Entry} &\bitbox{8}{Segment Left} & \bitbox[tlr]{16}{}\\
     \wordbox[lr]{1}{ Segment List[0](length based on SLoC type)  }\\
	     \wordbox[tlr]{3}{$\cdots$  }\\ 
	\wordbox[tlr]{1}{Segment List[N]} \\
		\bitbox[blr]{16}{}
		& \bitbox[tlr]{16}{} \\
	\wordbox[lrb]{2}{\dots\emph{optional} TLV} \\		
\end{bytefield}
	\caption{SRoU header}
\label{Fig:srou_hdr}
\end{figure}

\begin{itemize}
	
	\item\textbf{magic number:} 1Byte field with ALL ZERO. it used to distinguish packet in QUIC socket mode.
	\item\textbf{SRoU Length:} 1Byte, The total byte length of a SRoU header.
	\item\textbf{FT:} 2bits, defined flowid type, 0x0 = 32bits, 0x1 = 64bits, 0x2 = 96bits.
	\item\textbf{C:} C-bit, indicate the packet is encrypted .
	\item\textbf{F:} F-bit, indicate the packet encryption scope, 0x0 = SRoU header only , 0x1 = Full packet encryption.
	\item\textbf{T:} T-bit, indicate the packet need to send postcard telemetry to controller.
	\item\textbf{Protocol-ID:} 8bits, defined the inner packet protocol, 0x0 = SRoU OAM, 0x1 = IPv4, 0x2=IPv6
	\item\textbf{Source Address \& Port:} length based on Protocol-ID, OAM message does not contain this field.
	\item\textbf{SLoC Type:} 8bits field, 0x0 Reserved, 0x1 indicate the SLoC length is 48bits(IPv4+UDP port), 0x2 reserved for interworking with SRv6 with length equal 128bits, 0x3 used for Compressed Segment list.
	\item\textbf{SR Hdr Len:} R Header length, include the Header flags(4Bytes), Segment List and Optional TLV.
	\item\textbf{Last Entry:} contains the index(zero based), in the SLoC List, of the last element of the SLoC List.
	\item\textbf{Segments Left:}  8-bit unsigned integer.  Number of route segments	remaining, i.e., number of explicitly listed intermediate nodes still to be visited before reaching the final destination.
	\item\textbf{Segments List[0..n]:} SLoC store in each segment element.
	\item\textbf{Optional TLV:} 0x0 used for padding, 0x1 used for SR Integrity, 0x2 used for PathTelemetry.
	
\end{itemize}

\subsection{SRoU OAM}

SRoU OAM Message format defined as below:

\begin{figure}[h]
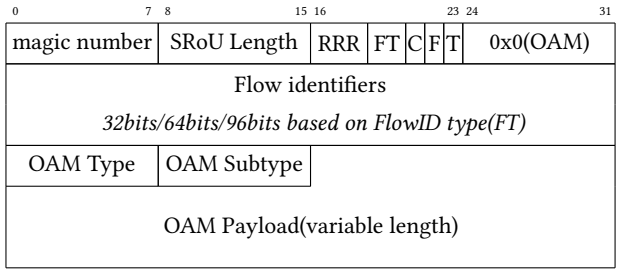

	\centering
	\begin{bytefield}[bitwidth=0.8em]{32}
		\bitheader{0,7,8,15,16,23,24,31} \\
		\bitbox{8}{magic number} & \bitbox{8}{SRoU Length} & \bitbox{3}{RRR} \bitbox{2}{FT} \bitbox{1}{C}\bitbox{1}{F}\bitbox{1}{T} &\bitbox{8}{0x0(OAM)}\\
		\wordbox[tlr]{1}{Flow identifiers} \\
		\wordbox[blr]{1}{\emph{ 32bits/64bits/96bits based on FlowID type(FT)}} \\
		\bitbox{8}{OAM Type} &\bitbox{8}{OAM Subtype} & \bitbox[tlr]{16}{}\\
		\wordbox[lbr]{2}{ OAM Payload(variable length)  }\\
	\end{bytefield}
	\caption{SRoU OAM header}
	\label{Fig:srou_oam_hdr}
\end{figure}

OAM type:=0x0 used for linkstate, subtype 0x0 used for Linkstate Request, subtype 0x1 used for Linkstate Response. the OAM Payload contains <seq(32bits)>, <timestamp(64bits)>,<received timestamp(64bits)>,<sender seq(32bits)> and <sender timestamp(64bits)>.

OAM type:=0x1 reserved for trace route, OAM type:=0x2 used for STUN service.

\subsection{Network programming}
Same as SRv6 Network Programming~\cite{srv6prog}, Ruta defined a virutal SLoC for network programming, IPv6 is same as RFC8986, Ruta under IPv4 based SLoC follows the following definition:

\begin{figure}[h]
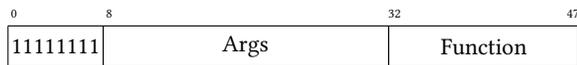

	\centering
	\begin{bytefield}[bitwidth=0.5em]{48}
		\bitheader{0,8,32,47} \\
		\bitbox{8}{11111111} & \bitbox{24}{Args} & \bitbox{16}{Function}\\
	\end{bytefield}
	\caption{SLoC for network function}
	\label{Fig:srou_hdr}
\end{figure}

We implement End.DT2U and End.DT4 in our prototype to provide VPN services.

%% file: sections/system.tex
\section{System Implementation}
\label{sec:system}

We implement Ruta prototype in native golang with nearly 9000 loc, cross compile could support multiple platform(x86/arm based linux, mips and arm based OpenWRT).
We also patch some codes to quic-go to support QUIC transport protocol over Ruta.

\subsection{Datacenter Spine-leaf deployment}

We deploy ruta with 2 linecard nodes as leaf ToR switch and 2 fabric nodes as spine, topology shown as below,demo code available in~\cite{ruta_demo}. 

\begin{figure}[h]
	\centering
	\includegraphics[scale=0.75]{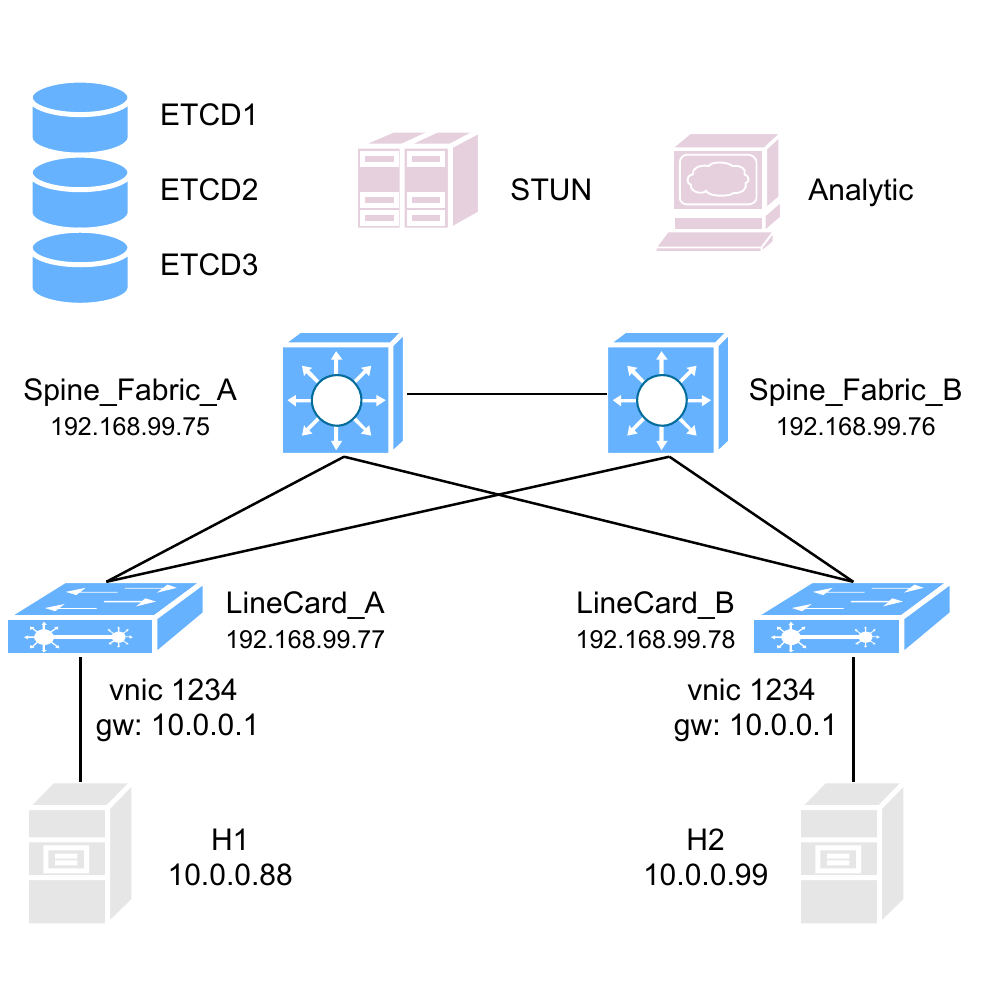}
	\caption{Spine Leaf deployment}
	\label{Fig:ruta_spine_leaf}
\end{figure}

ETCD cluster could be deployed in multiple region to provide better resilience, each ruta node could be run as etcd proxy mode to help other nodes use in-band communication to register to ETCD cluster.For instance Spine\_Fabric node could be etcd proxy for Linecard node in this deployment scenario. 

Linecard may have multiple uplinks, it could be represented and encoded in different colors and UDP ports in SLoC. The following chart shows the information retrieve from ETCD. Linecard will learn the MAC and IP address information from host and announce EVPN-Type2 route in ETCD, type-5 route were learned by local configuration or route redistribution  by other protocols.

\begin{figure}[h]
	\centering
	\includegraphics[scale=0.21]{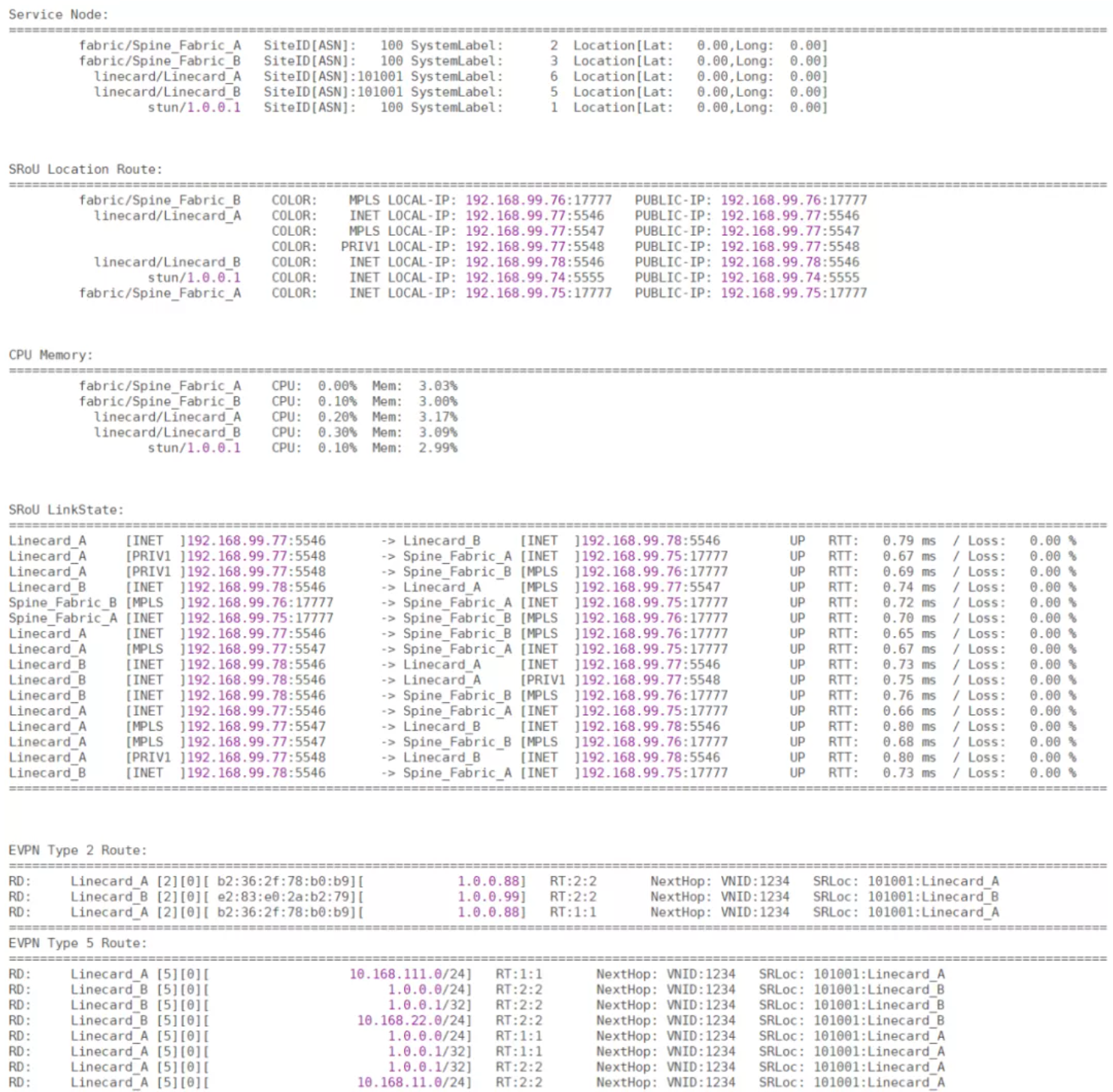}
	\caption{routing table dump from ETCD cluster}
	\label{Fig:ruta_spine_leaf1}
\end{figure}

FlexAlgo could be added in each linecard. In general case, H1->H2 communication could be encoded by Linecard1 with only one SLoC in SRoU Segment list, the interim Fabric just involved as basic IP forwarding function. 

\begin{figure}[h]
	\centering
	\begin{bytefield}[bitwidth=0.6em]{40}
		\wordbox[tlr]{1}{IP HDR } \\
		\wordbox[lr]{1}{ Src: 192.168.99.77(LC\_A) Dst: 192.168.99.78(LC\_B)} \\
		\wordbox{1}{UDP Src: 5547 Dst: 5546} \\
		\wordbox[lr]{1}{SRoU Hdr SL=1} \\		
		\wordbox[lrb]{1}{255.<vnid:1234>.End.DT2U} \\
		\wordbox[lr]{1}{Overlay Payload} \\		
		\wordbox[lbr]{2}{src:10.0.0.88, dst:10.0.0.99} \\
	\end{bytefield}
	\caption{SRoU Overlay Encapsulation}
	\label{Fig:spine_leaf_encap}
\end{figure}

The total SRoU header length is 24Bytes which is equal to SRv6 SRH, but it has IPv4 underlay which is more efficient than SRv6. Compare with VXLAN, the SRoU flowID header could be used as application aware tag or group based policy header, it could easily support traffic-engineering, however VXLAN need to use multiple NSH header instead. For instance, when network congestion, the Linecard could add more SLoC in segment list for traffic engineering. Spine-A will invoke the SRoU stack to relay packet to Linecard\_B.

\begin{figure}[h]
	\centering
	\begin{bytefield}[bitwidth=0.6em]{40}
		\wordbox[tlr]{1}{IP HDR } \\
		\wordbox[lr]{1}{ Src: 192.168.99.77(LC\_A) Dst: 192.168.99.75(Spine-A)} \\
		\wordbox{1}{UDP Src: 5547 Dst: 17777} \\
		\wordbox[lr]{1}{SRoU Hdr SL=2} \\		
		\wordbox[lr]{1}{255.<vnid:1234>.End.DT2U} \\
		\wordbox[lrb]{1}{192.168.99.78:5546(LC\_B SLoC)} \\
		\wordbox[lr]{1}{Overlay Payload} \\		
		\wordbox[lbr]{2}{src:10.0.0.88, dst:10.0.0.99} \\
	\end{bytefield}
	\caption{SRoU Traffic-Engineering Encapsulation}
	\label{Fig:spine_leaf_te}
\end{figure}

\subsection{Multi-cloud deployment}

Real time collaboration(RTC) like Cisco Webex, Zoom and Microsoft Teams are widely used during COVID-19, but the internet directly connection performance does not meet the application's requirement. Internet routing is based on BGP AS-PATH, congestion always observed between AS, can we build a pinhole between service providers without change any BGP routes? A multihoming ruta fabric node with SRoU encapsulation could easily implement packet relay between SP, meanwhile public cloud service providers always multi-homing with other traditional service providers and internet exchanges, deploy virtual machine on public cloud as Ruta fabric node could significantly improve RTC performance.

\begin{figure}[h]
	\centering
	\includegraphics[scale=0.18]{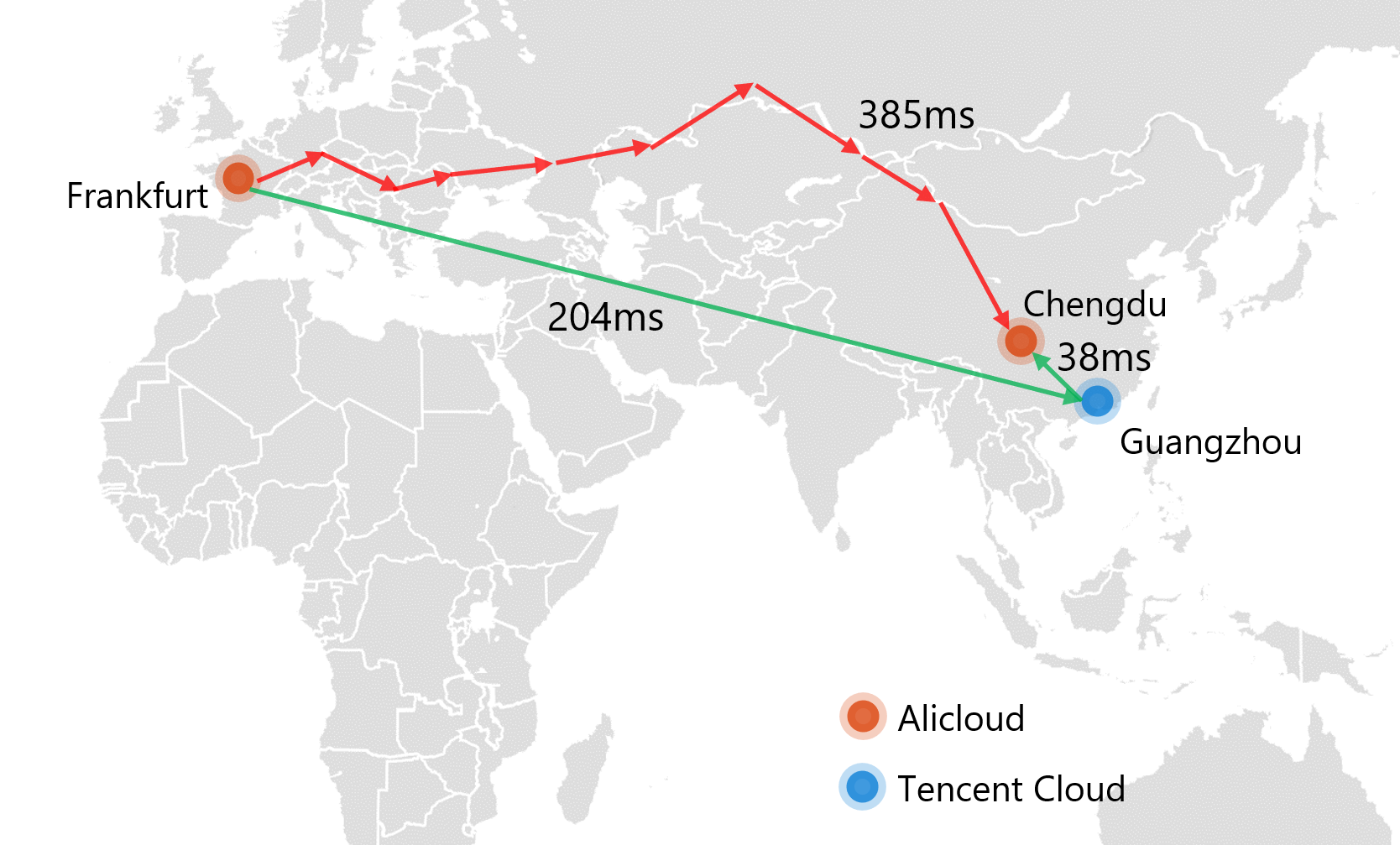}
	\caption{Multi cloud optimization}
	\label{Fig:multi_cloud}
\end{figure}

We need to emphasize multi-cloud deployment is useful,for instance, from Alicloud Frankfurt region to Alicloud Chengdu region, the directly internet connections over elastic IP over internet has 385 milliseconds latency, however if we relay it at Tencent Cloud guangzhou region will lower down the latency to 240 milliseconds.

Finally, we deploy 20 fabric nodes over multiple public cloud providers (Alicloud and Tencent) over the world, the result shows ruta could significantly reduce the latency and packet drop, It could provide \emph{nearly zero loss and almost less than 200ms latency to access anywhere in the world over internet} with maximum 4 segments, full result available in ~\cite{multi_cloud}. 

\subsection{Native Socket}

QUIC is a reliable and secured transport protocol in user-space, we implement SRoU function with quic-go to enhance the network programmability to provide traffic-engineering and multi-path forwarding capabilities over internet and directly traversal VPC.

\begin{figure}[h]
	\centering
	\includegraphics[scale=0.75]{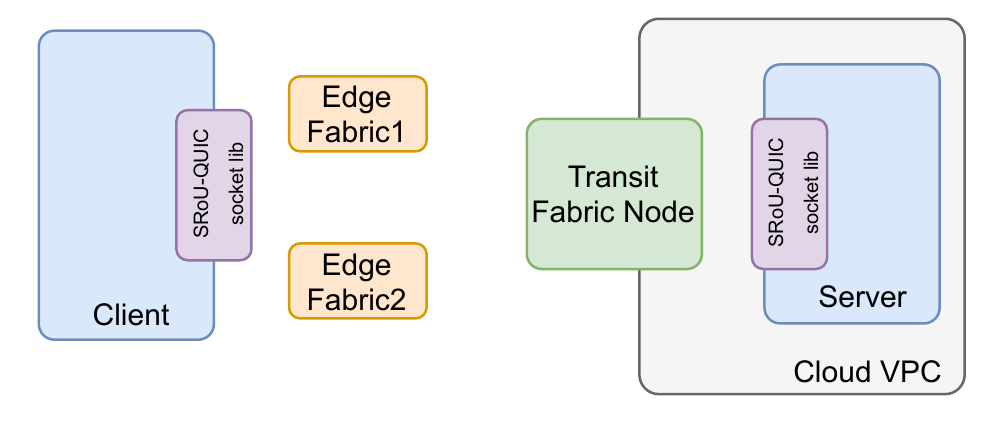}
	\caption{native socket lib}
	\label{Fig:native_socket}
\end{figure}

In native socket mode, server may register to Ruta ETCD to provide application awareness,especially in K8s deployment case, Ruta ETCD could directly sync the server's identity with K8s ETCD. Client does not required to register to ETCD, a simple DNS SRV record could be used to announce the edge\_fabric and transit\_fabric.

The client select edge fabric 1 could send the packet in the following format, consider the network security issue, Edge fabric1 could allocation some time based token for client, this token could be store in flowid field. 

\begin{table}[h!]
	\small
	\begin{tabular}{m{2.6cm}m{4.4cm}}
		\toprule
		\textbf{Field} & \textbf{Value}\\
		\hline
		IP Header & src: client\_ip dst: edge\_fabric1's ip  \\
		\hline
		SRoU Header &  \\
		magic number &  0 \\
		flowid & 32bits service token assigned by edge\_fabric1 \\
		source address & 0.0.0.0 \\
		source port & 0 \\
		segment list[0] & final server ip and port \\
		segment list[1] & transit fabric node \\
		\hline     	
		Inner Payload &  QUIC packet\\	
		\bottomrule
	\end{tabular}
	\caption{Service Locator(SLoC) Field}
	\label{tbl:SLoc_field}
\end{table}

client may not require to connect to STUN server for public address discovery, it simply fill ZERO in the SRoU source address and port field. The first hop fabric node will help copy the underlay source address inside the SRoU header for NAT-Traversal. when the server receive  the packet and find the UDP first 8 bits is ALLZERO, it will copy the internal SRoU header source address and port to Outer IP header and trim the SRoU to deliver the packet to application.

\subsection{Transparency VPC}
The native socket mode is not only support QUIC packets, but also support TCP and other UDP encapsulations. 

Existing VPC design requires cloud service provider manage the overlay routing table which makes some scale challenges for network devices ~\cite{sail_fish}. Many hybrid cloud solution require SDWAN or IPSec VPN as Gateway to bring traffic into the VPC, however it's very hard to inter working with overlay VPC routing table with existing private network, meanwhile each cloud provider has private micro-segmentation polices design, that may introduce significant efforts for interworking.

Ruta provide a cloud-native based approach to resolve the challenges, by enable Ruta linecard at each VPC and provide Segment Routing capability on overlay makes the VPC more transparency and the cloud provider could significantly reduce the overlay routing control and offload service gateway to each host or dedicated DPU. 

This dis-aggregation routing system could benefit edge cloud  and hybrid cloud deployment in the future.

\subsection{Large-scale deployment}
Some massive scale cloud service provider could use Ruta as distributed SDN solution. An Ruta system(with multiple LC and Fabric) could be treated as an single Linecard with multiple SLoC to join another Ruta system. Or multiple Ruta System could leverage on BGP-EVPN to share routing information as same as Inter-AS Option C MPLS VPN.

%% file: sections/conclusion.tex
\section{conclusion And Future Work}
\label{sec:conclusion}

Ruta is a dis-aggregated routing system based on cloud native approach, we develop a new K-V based control plane to decouple the configuration and management complexity and provide distributed deployment which perfectly resolve the centralized SDN controller challenge. Meanwhile we provide a Segment routing based transport layer, it provide \emph{nearly zero loss and almost less than 200ms latency to access anywhere in the world over internet}. We also enable native socket support for endpoint enable QUIC multi-path capabilities, even more we simplify the cloud VPC deployment and provide transparency VPC to support hybrid-cloud deployment.

Ruta is a first step towards cloud native datapath, we believe it cloud be used in more scenarios like NetDAM~\cite{fang2021netdam} for HPC , new container network interface and cloud-agnostic RPC framework in the future. we are exploring to enable Ruta for P4 based switch and DPU offload , server-less computing and datastreaming processing cases in the future.

%% file: sections/acknowledgment.tex
\section{acknowledgment}
\label{sec:acknowledgment}

It took a village to make Ruta possible, we thank Feng Cai, Yanhuan Mao, Yinghao Li,  Bin Shi, Yijen Wang, Xing Jiang, Yin Wang, Sam Gao support for this project.

%% file: main.bbl

\begin{thebibliography}{14}


\ifx \showCODEN    \undefined \def \showCODEN     #1{\unskip}     \fi
\ifx \showDOI      \undefined \def \showDOI       #1{#1}\fi
\ifx \showISBNx    \undefined \def \showISBNx     #1{\unskip}     \fi
\ifx \showISBNxiii \undefined \def \showISBNxiii  #1{\unskip}     \fi
\ifx \showISSN     \undefined \def \showISSN      #1{\unskip}     \fi
\ifx \showLCCN     \undefined \def \showLCCN      #1{\unskip}     \fi
\ifx \shownote     \undefined \def \shownote      #1{#1}          \fi
\ifx \showarticletitle \undefined \def \showarticletitle #1{#1}   \fi
\ifx \showURL      \undefined \def \showURL       {\relax}        \fi
\providecommand\bibfield[2]{#2}
\providecommand\bibinfo[2]{#2}
\providecommand\natexlab[1]{#1}
\providecommand\showeprint[2][]{arXiv:#2}

\bibitem[\protect\citeauthoryear{Brewer}{Brewer}{2021}]%
        {cap}
\bibfield{author}{\bibinfo{person}{Eric Brewer}.}
  \bibinfo{year}{2021}\natexlab{}.
\newblock \showarticletitle{{CAP theorem}}.
\newblock
  \bibinfo{howpublished}{{\url{https://en.wikipedia.org/wiki/CAP_theorem}}}.
\newblock  (\bibinfo{year}{2021}).
\newblock


\bibitem[\protect\citeauthoryear{C.~Filsfils}{C.~Filsfils}{2021}]%
        {srv6prog}
\bibfield{author}{\bibinfo{person}{J.~Leddy D. Voyer S. Matsushima Z. Li~Ed.
  C.~Filsfils, P.~Camarillo}.} \bibinfo{year}{2021}\natexlab{}.
\newblock \showarticletitle{{Segment Routing over IPv6 (SRv6) Network
  Programming}}.
\newblock
  \bibinfo{howpublished}{{\url{https://datatracker.ietf.org/doc/html/rfc8986}}}.
\newblock  (\bibinfo{year}{2021}).
\newblock


\bibitem[\protect\citeauthoryear{Cisco}{Cisco}{2021a}]%
        {ciscoaci}
\bibfield{author}{\bibinfo{person}{Cisco}.} \bibinfo{year}{2021}\natexlab{a}.
\newblock \showarticletitle{{Cisco Application Centric Infrastructure}}.
\newblock
  \bibinfo{howpublished}{{\url{https://www.cisco.com/c/en/us/solutions/data-center-virtualization/application-centric-infrastructure/index.html}}}.
\newblock  (\bibinfo{year}{2021}).
\newblock


\bibitem[\protect\citeauthoryear{Cisco}{Cisco}{2021b}]%
        {ciscodnac}
\bibfield{author}{\bibinfo{person}{Cisco}.} \bibinfo{year}{2021}\natexlab{b}.
\newblock \showarticletitle{{Cisco DNA Center Solution}}.
\newblock
  \bibinfo{howpublished}{{\url{https://www.cisco.com/c/en/us/products/cloud-systems-management/dna-center/index.html}}}.
\newblock  (\bibinfo{year}{2021}).
\newblock


\bibitem[\protect\citeauthoryear{cisco}{cisco}{2021}]%
        {lisp}
\bibfield{author}{\bibinfo{person}{cisco}.} \bibinfo{year}{2021}\natexlab{}.
\newblock \showarticletitle{{Locator/ID Separation Protocol}}.
\newblock
  \bibinfo{howpublished}{{\url{https://www.cisco.com/c/en/us/products/ios-nx-os-software/locator-id-separation-protocol-lisp/index.html}}}.
\newblock  (\bibinfo{year}{2021}).
\newblock


\bibitem[\protect\citeauthoryear{Fang and Peng}{Fang and Peng}{2021}]%
        {fang2021netdam}
\bibfield{author}{\bibinfo{person}{Kevin Fang} {and} \bibinfo{person}{David
  Peng}.} \bibinfo{year}{2021}\natexlab{}.
\newblock \showarticletitle{NetDAM: Network Direct Attached Memory with
  Programmable In-Memory Computing ISA}.
\newblock  (\bibinfo{year}{2021}).
\newblock
\showeprint[arxiv]{2110.14902}~[cs.DC]


\bibitem[\protect\citeauthoryear{J.Lemon}{J.Lemon}{2019}]%
        {vxlan_gpe_gbp}
\bibfield{author}{\bibinfo{person}{M.Smith~A.Isaac J.Lemon, F.Maino}.}
  \bibinfo{year}{2019}\natexlab{}.
\newblock \showarticletitle{{Group Policy Encoding with VXLAN-GPE and
  LISP-GPE}}.
\newblock
  \bibinfo{howpublished}{{\url{https://datatracker.ietf.org/doc/html/draft-lemon-vxlan-lisp-gpe-gbp-02}}}.
\newblock  (\bibinfo{year}{2019}).
\newblock


\bibitem[\protect\citeauthoryear{juniper}{juniper}{2021}]%
        {junipermist}
\bibfield{author}{\bibinfo{person}{juniper}.} \bibinfo{year}{2021}\natexlab{}.
\newblock \showarticletitle{{Juniper Mist AI solution}}.
\newblock
  \bibinfo{howpublished}{{\url{https://www.juniper.net/us/en/products/mist-ai.html}}}.
\newblock  (\bibinfo{year}{2021}).
\newblock


\bibitem[\protect\citeauthoryear{K.~Hedayat}{K.~Hedayat}{2008}]%
        {twamp}
\bibfield{author}{\bibinfo{person}{A.~Morton K. Yum J.~Babiarz K.~Hedayat,
  R.~Krzanowski}.} \bibinfo{year}{2008}\natexlab{}.
\newblock \showarticletitle{{Two-Way Active Measurement Protocol}}.
\newblock
  \bibinfo{howpublished}{{\url{https://datatracker.ietf.org/doc/html/rfc5357}}}.
\newblock  (\bibinfo{year}{2008}).
\newblock


\bibitem[\protect\citeauthoryear{K.Fang.}{K.Fang.}{2021a}]%
        {multi_cloud}
\bibfield{author}{\bibinfo{person}{K.Fang.}} \bibinfo{year}{2021}\natexlab{a}.
\newblock \showarticletitle{{Ruta Multicloud optimization result}}.
\newblock
  \bibinfo{howpublished}{{\url{https://github.com/zartbot/ruta_demo/blob/main/multicloud/log}}}.
\newblock  (\bibinfo{year}{2021}).
\newblock


\bibitem[\protect\citeauthoryear{K.Fang.}{K.Fang.}{2021b}]%
        {ruta_demo}
\bibfield{author}{\bibinfo{person}{K.Fang.}} \bibinfo{year}{2021}\natexlab{b}.
\newblock \showarticletitle{{Ruta Spine Leaf demo}}.
\newblock \bibinfo{howpublished}{{\url{https://github.com/zartbot/ruta_demo}}}.
\newblock  (\bibinfo{year}{2021}).
\newblock


\bibitem[\protect\citeauthoryear{Kreeger}{Kreeger}{2018}]%
        {vxlangbp}
\bibfield{author}{\bibinfo{person}{M.Smith~L. Kreeger}.}
  \bibinfo{year}{2018}\natexlab{}.
\newblock \showarticletitle{{VXLAN Group Policy Option}}.
\newblock
  \bibinfo{howpublished}{{\url{https://datatracker.ietf.org/doc/html/draft-smith-vxlan-group-policy-05}}}.
\newblock  (\bibinfo{year}{2018}).
\newblock


\bibitem[\protect\citeauthoryear{Pan, Yu, Jia, Pi, Xu, Qiao, Li, Liu, Lu, Lu,
  Song, Zhang, Huang, and Zhu}{Pan et~al\mbox{.}}{2021}]%
        {sail_fish}
\bibfield{author}{\bibinfo{person}{Tian Pan}, \bibinfo{person}{Nianbing Yu},
  \bibinfo{person}{Chenhao Jia}, \bibinfo{person}{Jianwen Pi},
  \bibinfo{person}{Liang Xu}, \bibinfo{person}{Yisong Qiao},
  \bibinfo{person}{Zhiguo Li}, \bibinfo{person}{Kun Liu}, \bibinfo{person}{Jie
  Lu}, \bibinfo{person}{Jianyuan Lu}, \bibinfo{person}{Enge Song},
  \bibinfo{person}{Jiao Zhang}, \bibinfo{person}{Tao Huang}, {and}
  \bibinfo{person}{Shunmin Zhu}.} \bibinfo{year}{2021}\natexlab{}.
\newblock \showarticletitle{Sailfish: Accelerating Cloud-Scale Multi-Tenant
  Multi-Service Gateways with Programmable Switches}.
\newblock  (\bibinfo{year}{2021}), \bibinfo{pages}{194–206}.
\newblock
\showISBNx{9781450383837}
\urldef\tempurl%
\url{https://doi.org/10.1145/3452296.3472889}
\showDOI{\tempurl}


\bibitem[\protect\citeauthoryear{wikipedia}{wikipedia}{2021}]%
        {fboutage}
\bibfield{author}{\bibinfo{person}{wikipedia}.}
  \bibinfo{year}{2021}\natexlab{}.
\newblock \showarticletitle{{Facebook 2021 Outage}}.
\newblock
  \bibinfo{howpublished}{{\url{https://en.wikipedia.org/wiki/2021_Facebook_outage}}}.
\newblock  (\bibinfo{year}{2021}).
\newblock


\end{thebibliography}
